\documentclass[twocolumn]{article}

\usepackage{float}
\usepackage{multicol}
\usepackage[font=small]{caption}
\usepackage{lipsum}
\usepackage{subfigure}
\usepackage{tabularx}
\usepackage{booktabs}
\usepackage{graphicx}
\usepackage{amsmath}
\usepackage{amsfonts}

\usepackage{latexsym}
\usepackage{mathrsfs}
\usepackage{geometry}
\usepackage{chngcntr}
\title{Making the case for causal dynamical triangulations}

\author{Joshua H. Cooperman \\ \emph{Institute for Mathematics, Astrophysics, and Particle Physics} \\ \emph{Radboud Universiteit Nijmegen, Heyendaalseweg 135, 6526 AJ Nijmegen, Nederland}}

\begin{document}




\twocolumn[
\begin{@twocolumnfalse}
\maketitle
\begin{abstract}
The aim of the causal dynamical triangulations approach is to define nonperturbatively a quantum theory of gravity as the continuum limit of a lattice-regularized model of dynamical geometry. 
My aim in this paper is to give a concise yet comprehensive, impartial yet personal presentation of the causal dynamical triangulations approach. 
\end{abstract}
\end{@twocolumnfalse}
]




\emph{Introduction}---Will the quantum theory of gravity prove to be yet another triumph of the paradigm of the quantum theory of fields? 
Any answer to this question---especially one based on current experimental knowledge and even one based on current theoretical knowledge---is extremely premature. Any attempt to construct a quantum theory of gravity as a quantum theory of fields---in light of this paradigm's spectacular successes---is clearly motivated. Well prior to these successes---indeed, 
concurrent with the 
development of quantum electrodynamics, our archetypal quantum theory of fields---were the first such attempts undertaken. Rosenfeld initiated a field-theoretic quantization of linearized general relativity \cite{LR1,LR2}, and Bronstein began to address the issues posed by the nonlinearity of general relativity for such a quantization \cite{MPB}. 


Through the work of myriad others, we have since learned that a field-theoretic quantization of general relativity cannot be precisely patterned on that of Maxwellian electrodynamics. 
As 't Hooft and Veltman first suggested and Goroff and Sagnotti then demonstrated, quantum general relativity is perturbatively nonrenormalizable \cite{GtH&MV,MHG&AS}. 
This technical result is by no means the death knell of a field-theoretic quantum theory of gravity. 
First of all, as Donoghue has stressed, this nonrenormalizability presents no obstruction to the derivation of universal predictions for sufficiently low energy quantum gravitational phenomena \cite{JD}. In other words, perturbative quantum general relativity constitutes a consistent effective theory whose predictions any quantum theory of gravity (with general relativity as its classical limit) must reproduce. 
Unfortunately, at least for experimental investigations, the quantum gravitational phenomena so predicted are exceedingly small effects as the inverse Planck energy sets their typical magnitudes. Potentially, nonpertubative quantum gravitational phenomena might be manifest at energy scales much smaller than the Planck energy.

Moreover, as Weinberg proposed, a quantum theory of fields can be asymptotically safe, implying nonperturbative renormalizability even in the absence of perturbative renormalizability 
\cite{SW}. We now possess examples of asymptotically safe quantum theories of fields, for example, certain nonlinear sigma models \cite{BHW&DK&AW}, and a considerable effort is currently underway to construct asymptotically safe quantum theories of gravity employing the exact (functional) renormalization group \cite{MN&MR}. 
Even if a quantum theory of gravity does not realize Weinberg's proposal, this theory might still prove effective 
over a considerable range of energies. 

Alternatively, one can attempt a field-theoretic quantization of a different classical theory of gravity. (Constructing a dynamically consistent and experimentally viable such theory is, however, quite nontrivial.) The resulting quantum theory of gravity might prove perturbatively renormalizable or asymptotically safe. This is the motivation behind, for instance, the currently fashionable Ho\v{r}ava-Lifshitz gravity \cite{PH1}.\footnote{Carlip discusses several other possible reactions to the perturbative nonrenormalizability of quantum general relativity \cite{SC}.}



A field-theoretic quantization of general relativity thus calls for the application of nonperturbative techniques.  Although exact renormalization group techniques have recently attracted much attention, 
lattice regularization techniques still remain the most well-established method for nonperturbative studies. The application of lattice regularization techniques to quantum theories of fields on Minkowski spacetime---exemplified by lattice quantum chromodynamics---has proved exceptionally successful \cite{QCD}. 
The philosophy behind lattice quantum gravity programs, of which causal dynamical triangulations is the latest incarnation, is precisely to follow as closely as possible the path traveled by lattice quantum chromodynamics. In this respect lattice quantum gravity programs are distinctly conservative approaches to the construction of quantum theories of gravity. 


Causal dynamical triangulations distinguishes itself in its assumption of one key hypothesis: all spacetimes contributing to the gravitational path integral must admit a global foliation by spacelike hypersurfaces all of the same topology. 
With this single hypothesis in place, treating the resulting theory in complete analogy to the treatment of lattice quantum chromodynamics, one obtains several promising initial results. 

In the following I aim to give a presentation of the causal dynamical triangulations approach that is 
accessible in its scope, candid in its appraisal, and independent in its perspective. 
I begin by formulating causal dynamical triangulations briefly but precisely, always emphasizing its patterning on lattice quantum chromodynamics. I next discuss the motivations for the hypothesis on which causal dynamical triangulations is based. I then relate the results that the causal dynamical triangulations approach has produced. I close by considering the most pressing questions facing the approach. 
I hope to portray causal dynamical triangulations as an approach conservative in both conception and intention, and I hope to make the case that causal dynamical triangulations merits further study. 
I refrain from comparisons with other approaches to the construction of quantum theories of gravity, obliging the causal dynamical triangulations approach to stand on its own merits. 

\emph{Formulation}---Consider a classical theory of gravity describing the dynamics of the spacetime metric tensor $\mathbf{g}$ according to the action $S_{\mathrm{cl}}[\mathbf{g}]$. 
A path integral quantization of this theory proceeds in complete analogy to that of a classical theory of fields on Minkowski spacetime. One first defines the quantum states or transition amplitudes $\mathscr{A}[\gamma]$ \emph{via} the formal path integral  
\begin{equation}\label{quantumstate}
\mathscr{A}[\gamma]=\int_{\mathbf{g}|_{\partial\mathscr{M}}=\gamma}\mathrm{d}\mu(\mathbf{g})\,e^{iS_{\mathrm{cl}}[\mathbf{g}]/\hbar}.
\end{equation}
The metric tensor $\gamma$ induced on the boundary $\partial\mathscr{M}$ of the spacetime manifold $\mathscr{M}$ by its metric tensor $\mathbf{g}$ specifies the quantum state $\mathscr{A}[\gamma]$. This classical boundary data might, for example, consist of the metric tensors 
of initial and final spacelike hypersurfaces. The integration extends over all physically distinct metric tensors $\mathbf{g}$ satisfying the boundary condition $\mathbf{g}|_{\partial\mathscr{M}}=\gamma$, each metric tensor $\mathbf{g}$ weighted by the product of the measure $\mathrm{d}\mu(\mathbf{g})$ and the exponential $e^{iS_{\mathrm{cl}}[\mathbf{g}]/\hbar}$. The vast majority of these metric tensors $\mathbf{g}$ are not solutions of the classical theory. One then computes the expectation value $\mathbb{E}\{\mathscr{O}[\mathbf{g}]\}$ of a physical observable $\mathscr{O}[\mathbf{g}]$ in the quantum state $\mathscr{A}[\gamma]$ \emph{via} the formal path integral
\begin{equation}\label{expectationvalue}
\mathbb{E}\{\mathscr{O}[\mathbf{g}]\}=\int_{\mathbf{g}|_{\partial\mathscr{M}}=\gamma}\mathrm{d}\mu(\mathbf{g})\,e^{iS_{\mathrm{cl}}[\mathbf{g}]/\hbar}\mathscr{O}[\mathbf{g}].
\end{equation}
Although my description of this quantization 
makes it appear quite straightforward, 
rigorously defining and computing the formal path integrals \eqref{quantumstate} and \eqref{expectationvalue} is notoriously difficult. 


Three principle difficulties arise in rendering well-defined the formal path integrals \eqref{quantumstate} and \eqref{expectationvalue}. Firstly, these path integrals are generically divergent. 
Contributing metric tensors $\mathbf{g}$ may possess excitations on arbitrarily small length scales. 
This problem is expected: it arises already in the quantization of classical theories of fields on Minkowski spacetime and indeed in the quantization of certain nonrelativistic systems. Its solution is well-understood in principle if not in practice: one requires a regularization of the formal path integrals \eqref{quantumstate} and \eqref{expectationvalue}, which one eventually attempts to remove through a renormalization process. 

Secondly, the measure $\mathrm{d}\mu(\mathbf{g})$ is not uniquely defined. At the very least one designs the measure $\mathrm{d}\mu(\mathbf{g})$ to eliminate all of the classical theory's redundancies of description, but considerable freedom still remains. Although typically not emphasized in presentations of the quantization of classical theories of fields on Minkowski spacetime, the problem arises in this setting as well, and its solution ultimately rests on matching the quantum theory to experiment. 

Thirdly, the set of all physically distinct metric tensors $\mathbf{g}$ is ambiguous. (This is a gross understatement. See, for instance, the discussion in \cite{JA&JJ}.) 
Any particular metric tensor $\bar{\mathbf{g}}$ is defined on a spacetime manifold $\bar{\mathscr{M}}$ of fixed topology $\bar{\mathscr{T}}$. Should the path integration thus extend only over metric tensors $\mathbf{g}$ defined on a spacetime manifold $\bar{\mathscr{M}}$ of a fixed topology $\bar{\mathscr{T}}$? Should the path integration also extend over all possible topologies $\mathscr{T}$ of the spacetime manifold $\mathscr{M}$? Or is there a physically motivated middle ground? This problem is essentially peculiar to the path integral quantization of classical theories of gravity. Its solution---or at least a potential solution---lies at the heart of causal dynamical triangulations. 

Any candidate quantum theory of gravity based on the formal path integrals \eqref{quantumstate} and \eqref{expectationvalue} must address these three difficulties. The causal dynamical triangulations approach prescribes particular solutions to each of these problems. Its formulation contains essentially just one key hypothesis, 
a proposed resolution of the third difficulty: the path integration extends over all metric tensors $\mathbf{g}$ defined on a spacetime manifold $\bar{\mathscr{M}}$ of the direct product form $\Sigma\times\mathrm{I}_{\mathrm{I\!R}}$, where $\Sigma$ is a fixed $d$-dimensional spatial manifold, and $\mathrm{I}_{\mathrm{I\!R}}$ is a real temporal interval. The topology $\bar{\mathscr{T}}$ of the spacetime manifold $\bar{\mathscr{M}}$ is thus fixed \emph{a priori}, and the spatial topology is forbidden from evolving. The adjective causal finds its origin in this hypothesis: a spacetime manifold $\bar{\mathscr{M}}$ of the form $\Sigma\times\mathrm{I}_{\mathrm{I\!R}}$, which clearly admits a global foliation by spacelike hypersurfaces all of the same topology, supports a strong notion of relativistic causality \cite{RG}. 

According to the causal dynamical triangulations approach, one should thus consider the formal path integral 
\begin{equation}\label{causalquantumstate}
\mathscr{A}_{\Sigma}[\gamma]=\int_{\substack{\bar{\mathscr{M}}\cong\Sigma\times\mathrm{I}_{\mathrm{I\!R}} \\ \mathbf{g}|_{\partial\bar{\mathscr{M}}}=\gamma}}\mathrm{d}\mu(\mathbf{g})\,e^{iS_{\mathrm{cl}}[\mathbf{g}]/\hbar}
\end{equation} 
instead of that of equation \eqref{quantumstate}. The corresponding restriction of the formal path integral \eqref{expectationvalue} is obvious. The integration now extends only over what I call causal spacetimes, those whose manifold $\bar{\mathscr{M}}$ is isomorphic to $\Sigma\times\mathrm{I}_{\mathrm{I\!R}}$.
Within the causal dynamical triangulations approach one thus defines a distinct quantum theory of gravity for each choice of spatial manifold $\Sigma$. 
One may certainly object to the key hypothesis on theoretical grounds---indeed, I myself have misgivings---but I do not know of any experimental evidence contradicting it. After finishing my exposition of the causal dynamical triangulations approach, I address the various motivations for the hypothesis. Ultimately, we will judge this hypothesis on the basis of the predictions to which it leads. 

Having assimilated this single hypothesis, the causal dynamical triangulations approach now proceeds 
in complete analogy to lattice quantum chromodynamics. To address the first difficulty---now the divergence of the formal path integral \eqref{causalquantumstate}---one invokes a lattice regularization. The chosen regularization---an adaptation of Regge calculus \cite{TR}---is designed to implement straightforwardly the approach's key hypothesis. 

\begin{figure}[h!]
\centering
\includegraphics[width=\linewidth]{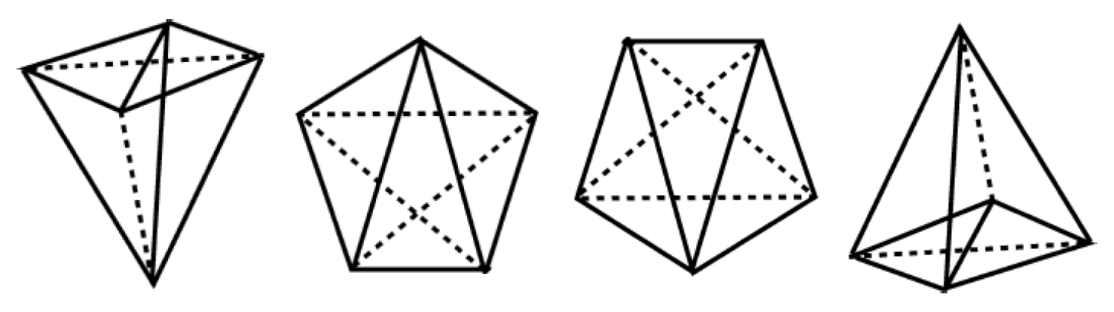}
\caption[left]{The four types of $4$-simplices employed in constructing $(3+1)$-dimensional causal triangulations: $(1,4)$ $4$-simplex, $(2,3)$ $4$-simplex, $(3,2)$ $4$-simplex, and $(4,1)$ $4$-simplex from left to right. The first number in the ordered pair indicates the number of vertices on an initial spacelike hypersurface, and the second number in the ordered pair indicates the number of vertices on the adjacent final spacelike hypersurface. The future timelike direction points from the bottom to the top of the page. I have taken these images from \cite{JA&JJ&RL2}.}
\label{4simplices}
\end{figure}

Continuous causal spacetimes 
are replaced by discrete causal triangulations.\footnote{When invoking this regularization, one implicitly assumes that the set of causal triangulations is in some sense sufficiently dense within the set of causal spacetimes.} 
A $(d+1)$-dimensional causal triangulation $\mathcal{T}_{c}$ is a Lorentzian simplicial manifold isomorphic to $\Sigma\times\mathrm{I}_{\mathrm{I\!R}}$ constructed from a large number $N_{d+1}$ of causal $(d+1)$-simplices. A causal $(d+1)$-simplex, of which there are $d+1$ types, is a particular simplicial piece of $(d+1)$-dimensional Minkowski spacetime. 
I depict the four types of causal $4$-simplices in figure \ref{4simplices}. Their spacelike edges have squared proper length $a^{2}$, establishing a lattice spacing $a$, and their timelike edges have squared proper length $-\alpha a^{2}$ for positive real parameter $\alpha$. 

The geometries of the causal $(d+1)$-simplices necessitate that, in their assembly into a causal triangulation $\mathcal{T}_{c}$, they come together in layers consistent with the manifold structure $\Sigma\times\mathrm{I}_{\mathrm{I\!R}}$. 
Each spatial layer is triangulated by regular spacelike $d$-simplices; 
adjacent spatial layers are connected by a single layer of timelike edges such that only the $d+1$ types of causal $(d+1)$-simplices are formed. 
The connectivities of all $N_{d+1}$ causal $(d+1)$-simplices comprising a causal triangulation $\mathcal{T}_{c}$, in combination with their edge length assignments, completely determine its metrical structure. While the interior of every causal $(d+1)$-simplex---a piece of Minkowski spacetime---has zero curvature, the junctions between different causal $(d+1)$-simplices---various types of subsimplices---carry nontrivial curvature. Most importantly, at least for the case considered below, the Ricci scalar curvature is concentrated on $(d+1-2)$-subsimplices.




A causal triangulation's spatial layers constitute the leaves of a distinguished global foliation by spacelike hypersurfaces all of the same topology. 
This distinguished foliation is not the only foliation of a causal triangulation---just as a spacetime manifold $\bar{\mathscr{M}}$ of the form $\Sigma\times\mathrm{I}_{\mathrm{I\!R}}$ does not possess a unique foliation---it is merely the foliation distinguished by a causal triangulation's skeleton. The distinguished foliation of a causal triangulation---indeed, the choice of lattice regularization---does constitute another postulate in the approach's formulation. Whether or not this postulate constitutes a further key hypothesis remains to be determined. 
One must ascertain whether or not the distinguished foliation persists in the continuum limit (if the latter exists).\footnote{If a continuum limit does not exist, then of course the distinguished foliation cannot be removed.} The continuum limit typically exhibits some degree of universality---insensitivity to certain features of the lattice regularization. Plausibly, the distinguished foliation is one such irrelevant detail. I comment further on this issue's status below. 



This lattice regularization results in the replacement of the formal path integral \eqref{causalquantumstate} by the concrete path sum
\begin{equation}\label{causalpathsum}
\mathcal{A}_{\Sigma}[\Gamma]=\sum_{\substack{\mathcal{T}_{c} \\ \mathcal{T}_{c}\cong\Sigma\times\mathrm{I}_{\mathrm{I\!R}} \\ \mathcal{T}_{c}|_{\partial\mathcal{T}_{c}}=\Gamma}}\mu(\mathcal{T}_{c})\,e^{i\mathcal{S}_{\mathrm{cl}}[\mathcal{T}_{c}]/\hbar}.
\end{equation}
The triangulation $\Gamma$ induced on the boundary $\partial\mathcal{T}_{c}$ of the causal triangulation $\mathcal{T}_{c}$ specifies the quantum state $\mathcal{A}_{\Sigma}[\Gamma]$. The summation extends over all causal triangulations $\mathcal{T}_{c}$ satisfying the boundary condition $\mathcal{T}_{c}|_{\partial\mathcal{T}_{c}}=\Gamma$, each causal triangulation $\mathcal{T}_{c}$ weighted by the product of the measure $\mu(\mathcal{T}_{c})$ and the exponential $e^{i\mathcal{S}_{cl}[\mathcal{T}_{c}]/\hbar}$. To address the second difficulty---now the nonuniqueness of the measure $\mu(\mathcal{T}_{c})$---one selects the minimal measure that eliminates redundancies of description: the inverse of the order of the automorphism group of the causal triangulation $\mathcal{T}_{c}$. The action $\mathcal{S}_{\mathrm{cl}}[\mathcal{T}_{c}]$ is the translation of the action $S_{\mathrm{cl}}[\mathbf{g}]$ into the Regge calculus of causal triangulations. One computes the expectation value $\mathbb{E}\{\mathcal{O}[\mathcal{T}_{c}]\}$ of a physical observable $\mathcal{O}[\mathcal{T}_{c}]$ in the quantum state $\mathcal{A}_{\Sigma}[\Gamma]$ by analogy to equation \eqref{expectationvalue}.


In principle, I have now completely specified a quantum theory of gravity---that resulting from a path integral quantization of the classical theory by the technique of causal dynamical triangulations. Two key questions immediately face this quantum theory of gravity. Firstly, as with any quantization of a classical theory, does there exist an appropriate limit in which the classical theory emerges from the quantum theory? I discuss this question's status below. 
If the correct classical limit emerges, then the quantum theory of gravity is at least viable. If the correct classical limit does not emerge, then the quantum theory of gravity is simply not viable. One should investigate the former theory's novel predictions while one should dispose of the latter theory. 

Secondly, as with any regularized quantum theory of fields, does there exist 
a continuum limit in which the regularization is removed in such a manner that physical quantities remain finite? 
I also discuss this question's status below. If a continuum limit exists, then the quantum theory of gravity is renormalizable. If a continuum limit does not exist, then the quantum theory of gravity is effective. The latter theory predicts its own breakdown on scales comparable to that of the lattice regularization; the former theory is ignorant, so to speak, of its eventual breakdown when novel dynamics on even smaller length scales asserts itself.


\emph{Investigation}---Analytically computing the path sum \eqref{causalpathsum} is currently an intractable problem in combinatorics beyond the simplest few cases in $1+1$ dimensions.\footnote{There is a sizable literature of analytical and numerical results on $(1+1)$-dimensional causal dynamical triangulations. See, for instance, \cite{JA&AG&JJ&RL3}.} One thus turns to numerical techniques. Continuing to follow the path traveled by lattice quantum chromodynamics, one would like to employ Markov chain Monte Carlo methods to generate an ensemble of causal triangulations representative of those contributing to the path sum \eqref{causalpathsum}. Despite notable progress in expanding their applicability \cite{KNA&TA&JN}, Markov chain Monte Carlo methods still essentially require that the configurations to be generated are weighted by real rather than complex numbers. In the setting of lattice quantum chromodynamics, one basically addresses this issue by Wick rotation. If one could perform a Wick rotation of each causal triangulation from the Lorentzian to the Euclidean sector, then one could employ Markov chain Monte Carlo methods. 

A causal triangulation's structure---specifically, its global foliability by spacelike hypersurfaces---allows for a well-defined Wick rotation. The parameter $\alpha$, which dictates the relative scaling of spacelike to timelike squared edge lengths, holds the key to this Wick rotation. By analytically continuing $\alpha$ to $-\alpha$ through the lower half complex plane, one brings each causal triangulation from the Lorentzian to the Euclidean sector. One thus transforms the path sum \eqref{causalpathsum} into the partition function
\begin{equation}\label{causalpartitionfunction}
\mathcal{Z}_{\Sigma}[\Gamma]=\sum_{\substack{\mathcal{T}_{c} \\ \mathcal{T}_{c}\cong\Sigma\times\mathrm{I}_{\mathrm{I\!R}} \\ \mathcal{T}_{c}|_{\partial\mathcal{T}_{c}}=\Gamma}}\mu(\mathcal{T}_{c})\,e^{-\mathcal{S}_{\mathrm{cl}}^{(\mathrm{E})}[\mathcal{T}_{c}]/\hbar}
\end{equation}
for the real-valued Euclidean action $\mathcal{S}_{\mathrm{cl}}^{(\mathrm{E})}[\mathcal{T}_{c}]$.\footnote{The transfer matrix associated with the partition function \eqref{causalpartitionfunction} is essentially reflection positive, indicating that the quantum theory so defined is unitary \cite{JA&JJ&RL2}.}

The partition function \eqref{causalpartitionfunction} defines a statistical mechanical model to which one brings to bear standard techniques for its study. One first runs Markov chain Monte Carlo simulations of causal triangulations representative of those contributing to the partition function \eqref{causalpartitionfunction}. These simulations employ a standard Metropolis algorithm based on the well-known Pachner moves adapted to causal triangulations \cite{JA&JJ&RL2}.\footnote{Technically, no one has yet proved that the Pachner moves for $(2+1)$- and $(3+1)$-dimensional causal triangulations are rigorously ergodic.} One thus generates ensembles of causal triangulations characterized by the number $T$ of the distinguished foliation's leaves, the number $N_{d+1}$ of causal $(d+1)$-simplices, and the bare couplings of the action $\mathcal{S}_{\mathrm{cl}}^{(\mathrm{E})}[\mathcal{T}_{c}]$.\footnote{To simulate causal triangulations numerically, one must restrict to finite $T$ and $N_{d+1}$. Fixing the values of $T$ and $N_{d+1}$ 
within a simulation is a convenient choice particularly well-suited to finite size scaling analysis. 
The partition function for fixed $T$ and $N_{d+1}$ is related by a Legendre transform to the partition function \eqref{causalpartitionfunction}.} One then analyzes these ensembles by performing numerical measurements of physical observables $\mathcal{O}[\mathcal{T}_{c}]$, estimating their expectation values $\mathbb{E}\{\mathcal{O}[\mathcal{T}_{c}]\}$ in the quantum state $\mathcal{Z}_{\Sigma}[\Gamma]$ by their averages $\langle\mathcal{O}[\mathcal{T}_{c}]\rangle$ over the ensemble. To remove the influence of the finite values of $T$ and $N_{d+1}$ on the ensemble averages $\langle\mathcal{O}[\mathcal{T}_{c}]\rangle$, one employs finite size scaling techniques.

Since one only introduced causal triangulations to regularize the formal path integral \eqref{causalquantumstate}, one would like to investigate the possibility of a continuum limit in which one could take the lattice spacing to zero. The existence of a continuum limit is contingent on the presence of an ultraviolet fixed point, and the existence of an ultraviolet fixed point is contingent on the presence of a second order phase transition. One thus explores the phase structure of the partition function \eqref{causalpartitionfunction} in search of such a transition, distinguishing phases by the respective values of certain physical observables---order parameters. If a second order phase transition exists, then 
one performs a renormalization group analysis to probe the presence of an ultraviolet fixed point along this transition. 

\emph{Motivation}---The hypothesis on which causal dynamical triangulations is founded finds motivations \emph{a priori} and \emph{a posteriori}. Its original inspiration came from the failures of the quantum Regge calculus and Euclidean dynamical triangulations approaches to produce physically sound quantum theories of gravity. 
Both of these lattice quantum gravity programs took as their starting point the formal partition function 
\begin{equation}\label{euclideanquantumstate}
\mathscr{Z}[\gamma]=\int_{\mathbf{g}|_{\partial\mathscr{M}}=\gamma}\mathrm{d}\mu(\mathbf{g})\,e^{-S_{\mathrm{cl}}^{(\mathrm{E})}[\mathbf{g}]/\hbar}
\end{equation}
for the Euclidean action $S_{\mathrm{cl}}^{(\mathrm{E})}[\mathbf{g}]$. After introducing their respective lattice regularizations of the partition function \eqref{euclideanquantumstate}, these approaches found neither phases of quantum geometry exhibiting classical features on sufficiently large length scales nor second order phase transitions between the phases present \cite{RL}. 
These findings pointed to the necessity of accounting for spacetime's Lorentzian causal structure within a path integral quantization. 

As an initial step in this direction, Ambj\o rn and Loll formulated causal dynamical triangulations in $1+1$ dimensions \cite{JA&RL}. They had as their primary purpose the lattice regularization of the formal path integral \eqref{quantumstate} directly within the Lorentzian sector. Acknowledging the severe ambiguity in the instruction to integrate over all metric tensors and anticipating the necessity of employing numerical techniques, they proposed the regularization by causal triangulations. 

For the Einstein-Hilbert action of general relativity---just the cosmological constant interaction in $1+1$ dimensions---Ambj\o rn, Correia, Kristjansen, and Loll then demonstrated a succinct relationship between Euclidean dynamical triangulations and (Wick-rotated) causal dynamical triangulations: by completely integrating baby universes out of the former, one obtains the latter, and by systematically integrating baby universes into the latter, one obtains the former \cite{ACJL}. (The absence of spatial topology change prevents the formation of baby universes in causal dynamical triangulations.) Furthermore, studies of Euclidean dynamical triangulations indicated that triangulations replete with baby universes dominate its partition function, implicating these configurations as the source of this approach's aphysicality \cite{SJ&SDM}. 
These results motivated Ambj\o rn, Jurkiewicz, and Loll to formulate causal dynamical triangulations in $2+1$ and $3+1$ dimensions \cite{JA&JJ&RL2,JA&JJ&RL1}. 






While these considerations make for a compelling retrospective narrative, 
one might still worry about the potential physical consequences of the approach's key hypothesis. 
The exclusion of spatial topology change from a quantum theory of gravity is possibly cause for concern. Many solutions of general relativity---the desired classical limit---allow for spatial topology change, but this feature 
may prove physically irrelevant. Certainly, it plays no role within the settings in which we have tested general relativity. Even if the quantum theory of gravity permits spatial topology change, quantum theories of gravity defined by causal dynamical triangulations might nevertheless accurately describe the theory's sectors in which spatial topology change does not occur. Unless spatial topology change is inseparable from the quantum dynamics of gravity, which could be the case, causal dynamical triangulations could well be of use. 

The presence of a distinguished foliation in the definition of a quantum theory of gravity is definitely cause for concern. Only certain solutions of general relativity 
possess such a foliation, and there exist stringent experimental constraints on the degree to which such a foliation may be manifest \cite{DM}. As I observed above, as an aspect of the lattice regularization, the distinguished foliation only has physical import if it persists in some form in the (hypothetical) continuum limit. What comes of the distinguished foliation in this limit is thus the crucial question. 

Ambj\o rn, Glaser, Sato, and Watabiki recently answered this question for the causal dynamical triangulations of $(1+1)$-dimensional general relativity \cite{JA&LG&YS&YW}. This theory's continuum limit is quantum projectable Ho\v{r}ava-Lifshitz gravity, itself a theory with a distinguished foliation. Although there exist several suggestive similarities between Ho\v{r}ava-Lifshitz gravity and the causal dynamical triangulations of $(2+1)$- and $(3+1)$-dimensional general relativity \cite{CDTandHL}, there is scant evidence for the distinguished foliation's persistence in these theories' continuum limits. Indeed, Jordan and Loll recently considered a generalization of causal dynamical triangulations in which the distinguished foliation is absent. Instead, only a local notion of causality based on the light cone structure at each vertex is enforced. Applying this new quantization to $(2+1)$-dimensional general relativity, these authors found preliminarily evidence that standard and generalized causal dynamical triangulations belong to the same universality class \cite{SJ&RL1,SJ&RL2}. 

Of course, the primary motivation for the approach's key hypothesis comes from the promising results that causal dynamical triangulations has produced. 

\emph{Results}---In the physically relevant setting of $3+1$ dimensions, the partition function \eqref{causalpartitionfunction} has only been studied numerically for the Einstein-Hilbert action of general relativity,
\begin{equation}\label{EHaction}
S_{\mathrm{cl}}[\mathbf{g}]=\frac{1}{16\pi G}\int_{\bar{\mathscr{M}}}\mathrm{d}^{4}x\sqrt{-g}\,(R-2\Lambda),
\end{equation} 
for a spacetime manifold $\bar{\mathscr{M}}$ of the form $\mathrm{S}^{3}\times\mathrm{S}^{1}$, the direct product of a spatial $3$-sphere and a temporal $1$-sphere. The real interval $\mathrm{I}_{\mathrm{I\!R}}$ is periodically identified for ease of numerical implementation. As I explain below, this identification does not influence the results. In this case one may cast the action $\mathcal{S}_{\mathrm{cl}}^{(\mathrm{E})}[\mathcal{T}_{c}]$ into the form
\begin{eqnarray}\label{EReggeaction4}
\mathcal{S}_{\mathrm{cl}}^{(\mathrm{E})}[\mathcal{T}_{c}]&=&(-\kappa_{0}+6\Delta)N_{0}+\Delta N_{4}^{(1,4)}\nonumber\\ &&\quad+\Delta N_{4}^{(4,1)}+(\Delta+\kappa_{4})N_{4}.
\end{eqnarray}
The bare couplings $\kappa_{0}$, $\Delta$, and $\kappa_{4}$ are specific functions of $G/a^{2}$, $\Lambda a^{2}$, and $\alpha$. $N_{0}$ denotes the number of vertices, $N_{4}^{(1,4)}$ denotes the number of $(1,4)$ $4$-simplices, and $N_{4}^{(4,1)}$ denotes the number of $(4,1)$ $4$-simplices. For this choice of spacetime topology, the partition function \eqref{causalpartitionfunction} essentially defines the ground state of quantum geometry in that there is no classical boundary data acting as an external source.

The partition function \eqref{causalpartitionfunction} for the action \eqref{EReggeaction4} exhibits three phases of quantum geometry: the decoupled phase A, the crumpled phase B, and the physical phase C. There is now compelling evidence that the AC phase transition is of first order and that the BC phase transition is of second order \cite{JA&SJ&JJ&RL2}. 
This last fact raises the possibility of rigorously defining the continuum limit of this quantum theory of gravity. Whether or not there are any fixed points---either infrared or ultraviolet---along this second order phase transition is not yet known, but efforts are currently underway to make this determination \cite{JA&AG&JJ&AK&RL,JHC1}. 

Studies of the quantum geometry within phase C have furnished convincing evidence that causal dynamical triangulations yields predictions in accord with general relativity in the classical limit, in accord with the quantum theory of fields on solutions of general relativity in the semiclassical limit, and of a novel phenomenon in the quantum-mechanical regime. These results are of precisely the character that one wants a quantum theory of gravity to produce: they agree with the classical theory in the classical limit, they agree with a perturbative quantization of the classical theory in the semiclassical limit, and they evidence novel phenomenology in the quantum-mechanical regime. 

\begin{figure}[h!]
\centering
\includegraphics[width=\linewidth]{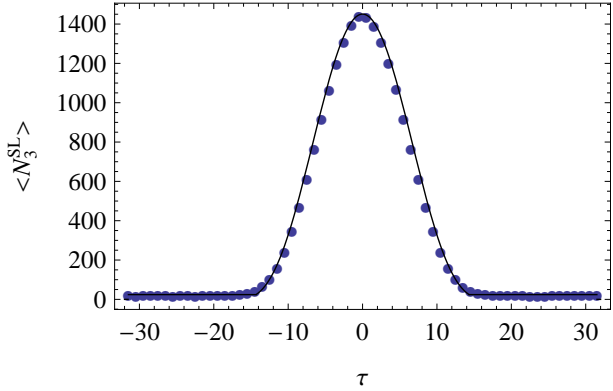}
\caption[left]{Fit of the spatial $3$-volume as a function of global time of Euclidean de Sitter space (black line) to the ensemble average number $\langle N_{3}^{\mathrm{SL}}\rangle$ of spacelike $3$-simplices as a function of the discrete time coordinate $\tau$ (dark blue dots) for an ensemble within phase C characterized by $T=64$, $N_{4}=81920$, $\kappa_{0}=2.0$, $\kappa_{4}=0.2$, and $\Delta=0.4$.}
\label{volprofiles}
\end{figure}

The first result stems from numerical measurements of $N_{3}^{\mathrm{SL}}(\tau)$, the number $N_{3}^{\mathrm{SL}}$ of regular spacelike $3$-simplices---equilateral tetrahedra---as a function of the discrete time coordinate $\tau$ labeling the distinguished foliation's $T$ leaves. One may conceive of $N_{3}^{\mathrm{SL}}(\tau)$ as the discrete time evolution of the discrete spatial $3$-volume.\footnote{The discrete function $N_{3}^{\mathrm{SL}}(\tau)$ is not necessarily a physical observable because of its dependence on a particular foliation's time coordinate; nevertheless, it does contain physical information.} Since $N_{3}^{\mathrm{SL}}(\tau)$ quantifies a property of each entire leaf, its measurement probes the quantum geometry on large length scales. In figure \ref{volprofiles} I depict the ensemble average $\langle N_{3}^{\mathrm{SL}}(\tau)\rangle$ for a typical ensemble of causal triangulations within phase C. Although the real interval $\mathrm{I}_{\mathrm{I\!R}}$ is periodically identified, the dynamics encoded in the partition function \eqref{causalpartitionfunction} restrict the quantum geometry to a portion of the $1$-sphere $\mathrm{S}^{1}$.

Given the regularity of $\langle N_{3}^{\mathrm{SL}}(\tau)\rangle$, one might suspect that an effective action $\mathcal{S}_{\mathrm{eff}}^{(\mathrm{E})}[N_{3}^{\mathrm{SL}}]$ governs the dynamics of $N_{3}^{\mathrm{SL}}(\tau)$. Numerical measurements---most accurately through those of the discrete time transfer matrix---reveal the action $\mathcal{S}_{\mathrm{eff}}^{(\mathrm{E})}[N_{3}^{\mathrm{SL}}]$ to have the form of a naive discretization of the Euclidean Einstein-Hilbert action within a minisuperspace truncation \cite{JA&JJ&RL4,JA&JJ&RL5,JA&JJ&RL6,JA&AG&JJ&RL2,JA&AG&JJ&RL&JGS&TT,JA&JGS&AG&JJ}. Crucially, though, the action $\mathcal{S}_{\mathrm{eff}}^{(\mathrm{E})}[N_{3}^{\mathrm{SL}}]$ differs by an overall negative sign. The Euclidean Einstein-Hilbert action suffers from not being bounded from below because its kinetic term has the wrong sign, a property shared by the action \eqref{EReggeaction4}. The immunity of the action $\mathcal{S}_{\mathrm{eff}}^{(\mathrm{E})}[N_{3}^{\mathrm{SL}}]$ to this disease most likely arises from the combinatorics of causal triangulations. If causal triangulations having values of the action \eqref{EReggeaction4} well-bounded from below vastly outnumber causal triangulations having values of the action \eqref{EReggeaction4} not well-bounded from below, then such combinatorics can counterbalance the weighting by the action \eqref{EReggeaction4} in the partition function \eqref{causalpartitionfunction} \cite{JA&AG&JJ&RL3}. 

The naive continuum limit of the action $\mathcal{S}_{\mathrm{eff}}^{(\mathrm{E})}[N_{3}^{\mathrm{SL}}]$ 
has as its maximally symmetric solution Euclidean de Sitter space. The shape of $\langle N_{3}^{\mathrm{SL}}(\tau)\rangle$ is very accurately modeled as the shape of Euclidean de Sitter space when the former's discrete time coordinate is identified with the latter's global time coordinate \cite{JA&JJ&RL4,JA&JJ&RL5,JA&JJ&RL6,JA&AG&JJ&RL1,JA&AG&JJ&RL2,JA&AG&JJ&RL&JGS&TT,JHC&JMM,CA&SJC&JHC&PH&RKK&PZ}. In figure \ref{volprofiles} I plot a fit to $\langle N_{3}^{\mathrm{SL}}(\tau)\rangle$ of the global time evolution of the spatial $3$-volume of Euclidean de Sitter space. On its largest length scales the ensemble average quantum geometry within phase C thus closely approximates that of Euclidean de Sitter space. 

\begin{figure}[h!]
\centering
\includegraphics[width=\linewidth]{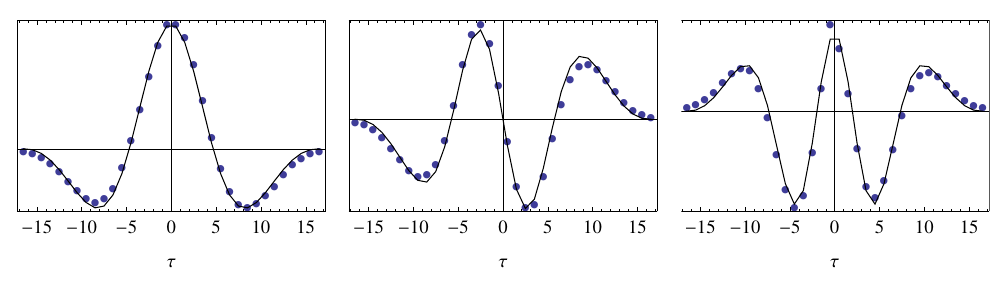}
\caption[left]{Fit of the model described in the text (black line) to first three eigenvectors of the ensemble average covariance $\langle \delta N_{3}^{\mathrm{SL}}(\tau)\delta N_{3}^{\mathrm{SL}}(\tau')\rangle$ (dark blue dots) for an ensemble within phase C characterized by $T=64$, $N_{4}=81920$, $\kappa_{0}=2.0$, $\kappa_{4}=0.2$, and $\Delta=0.4$.}
\label{eigenvectorfits}
\end{figure}
\begin{figure}[h!]
\centering
\includegraphics[width=\linewidth]{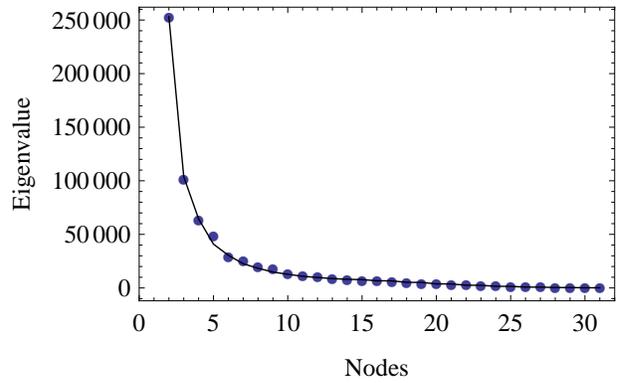}
\caption[left]{Fit of the model described in the text (black line) to the eigenvalues of the ensemble average covariance $\langle \delta N_{3}^{\mathrm{SL}}(\tau)\delta N_{3}^{\mathrm{SL}}(\tau')\rangle$ (dark blue dots) for an ensemble within phase C characterized by $T=64$, $N_{4}=81920$, $\kappa_{0}=2.0$, $\kappa_{4}=0.2$, and $\Delta=0.4$.}
\label{eigenvaluefit}
\end{figure}

The second result---that concerning the semiclassical limit---stems from numerical measurements of $\delta N_{3}^{\mathrm{SL}}(\tau)\delta N_{3}^{\mathrm{SL}}(\tau')$,  the covariance of deviations $\delta N_{3}^{\mathrm{SL}}(\tau)$ from the ensemble average $\langle N_{3}^{\mathrm{SL}}(\tau)\rangle$. One may conceive of $\delta N_{3}^{\mathrm{SL}}(\tau)\delta N_{3}^{\mathrm{SL}}(\tau')$ as the discrete time propagator or connected $2$-point function of fluctuations in the discrete spatial $3$-volume. The ensemble average covariance $\langle \delta N_{3}^{\mathrm{SL}}(\tau)\delta N_{3}^{\mathrm{SL}}(\tau')\rangle$ is well-modeled as follows \cite{JA&AG&JJ&RL2}. Consider linear perturbations of the metric tensor $\mathbf{g}$ propagating on the background of Euclidean de Sitter space restricted to depend only on the global time coordinate. The naive continuum limit of the action $\mathcal{S}_{\mathrm{eff}}^{(\mathrm{E})}[N_{3}^{\mathrm{SL}}]$ expanded to second order in these perturbations governs their dynamics. Perturbatively quantizing this classical theory, one can compute the naive continuum limit of the covariance $\langle \delta N_{3}^{\mathrm{SL}}(\tau)\delta N_{3}^{\mathrm{SL}}(\tau')\rangle$ by standard techniques. For a typical ensemble of causal triangulations within phase C, in figure \ref{eigenvectorfits} I plot fits to the first three eigenvectors of $\langle \delta N_{3}^{\mathrm{SL}}(\tau)\delta N_{3}^{\mathrm{SL}}(\tau')\rangle$ of the first three eigenvectors predicted by this model, 
and in figure \ref{eigenvaluefit} I plot a fit to the eigenvalues of $\langle \delta N_{3}^{\mathrm{SL}}(\tau)\delta N_{3}^{\mathrm{SL}}(\tau')\rangle$ of the eigenvalues predicted by this model.
As expected, the semiclassical model provides a better description of those eigenvectors (and their associated eigenvalues) with fewer nodes: the number of nodes of one of these excitations of the quantum geometry is inversely related to its characteristic length scale. 

Just how consequential are these findings? After all, the Einstein-Hilbert action \eqref{EHaction} contains the two most relevant interactions on large length scales, the classical action's extrema often dominate the path integral, and the ground state is typically the most symmetric quantum state. Euclidean de Sitter space is, however, only a saddle point of the Euclidean Einstein-Hilbert action, so its dominance of the partition function \eqref{causalpartitionfunction} for the action \eqref{EReggeaction4} is a nontrivial---and presumably nonperturbative---effect \cite{JA&AG&JJ&RL3}. 

\begin{figure}[h!]
\centering
\includegraphics[width=\linewidth]{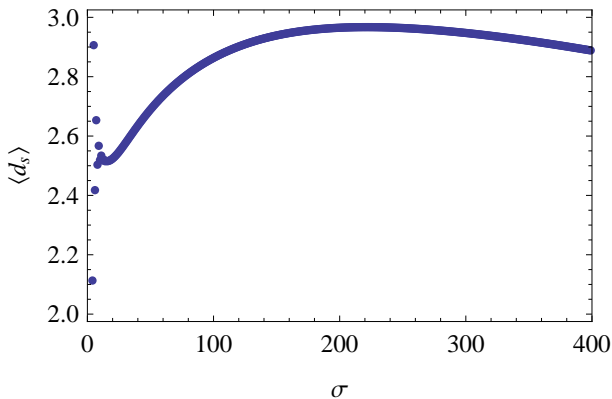}
\caption[left]{The ensemble average spectral dimension $\langle d_{s}\rangle$ as a function of diffusion time $\sigma$ for an ensemble within phase C characterized by $T=64$, $N_{3}=61440$, $k_{0}=1.0$, and $k_{3}=0.75$.} 
\label{specdim}
\end{figure}

The third result---that concerning the quantum-mechanical regime---stems from numerical measurements of $d_{s}(\sigma)$, the spectral dimension $d_{s}$ as a function of the diffusion time $\sigma$. The spectral dimension measures the effective dimensionality of a space as witnessed by a diffusing test random walker. The diffusion time, which just counts the number of steps in the diffusion process, serves as a proxy for the length scale probed. For a typical ensemble of causal triangulations within phase C, the ensemble average spectral dimension $\langle d_{s}(\sigma)\rangle$ has the characteristic form depicted in figure \ref{specdim}. (Note that I measured the spectral dimension of an ensemble of $(2+1)$-dimensional causal triangulations.)
On sufficiently small length scales the ensemble average spectral dimension has a value of approximately $2$ \cite{JA&JJ&RL7,JA&JJ&RL6,RK}.\footnote{The falloff towards a value of $2$ is typically more pronounced for a larger number $N_{d+1}$ of causal $(d+1)$-simplices.} Since the spectral dimension of a $(d+1)$-dimensional Euclidean manifold strictly approaches the topological value of $d+1$ as the diffusion time approaches zero, the dynamical reduction in the spectral dimension appears to be quantum-mechanical in origin. 

The ensemble average spectral dimension gradually increases with the diffusion time to peak at approximately the topological dimension of $4$---or rather $3$ for the case of figure \ref{specdim}---on intermediate length scales \cite{JA&JJ&RL7,JA&JJ&RL6,RK}. This constitutes strong evidence for classicality of the ensemble average quantum geometry on these scales. Numerical measurements of the Hausdorff dimension further support this finding \cite{JA&JJ&RL6,RK}. There are also indications that the ensemble average spectral dimension on sufficiently large length scales has the shape expected for Euclidean de Sitter space \cite{DB&JH}. 

While these are the primary results of the causal dynamical triangulations approach, several secondary results deserve brief mention. Applying causal dynamical triangulations to $(2+1)$-dimensional general relativity, several authors have demonstrated that the three primary results obtain in this case as well \cite{DB&JH,JA&JJ&RL3,JHC&JMM,JHC&KL&JMM}. The phase structure of the partition function \eqref{causalpartitionfunction} is different, however, as the crumpled phase B is absent \cite{RK,JA&JJ&RL3}. Cooperman and Miller studied this case for a spacetime manifold $\bar{\mathscr{M}}$ of the form $\mathrm{S}^{2}\times\mathrm{I}_{\mathrm{I\!R}}$ \cite{JHC&JMM,JHC&KL&JMM}, and Budd and Loll studied this case for a spacetime manifold $\bar{\mathscr{M}}$ of the form $\mathrm{T}^{2}\times\mathrm{I}_{\mathrm{I\!R}}$ \cite{TB&RL}. Also, Sachs numerically measured the spectral decomposition of the distinguished foliation's leaves in this case, finding their ensemble average quantum geometries to be exceedingly round \cite{MKS}. Recently, Cooperman ascertained the degree to which the ensemble average quantum geometry is homogeneous, finding evidence for inhomogeneity on sufficiently small length scales and for homogeneity on sufficiently large length scales \cite{JHC2}. Anderson \emph{et al} applied the causal dynamical triangulations approach to the quantization of $(2+1)$-dimensional projectable Ho\v{r}ava-Lifshitz gravity for a spacetime manifold $\bar{\mathscr{M}}$ of the form $\mathrm{S}^{2}\times\mathrm{S}^{1}$ \cite{CA&SJC&JHC&PH&RKK&PZ}. This last study provides the first indications that the quantum geometry of phase C, which now occurs within an enlarged phase diagram, is robust under the addition of higher order interactions to the classical action.

All of the results that I have chronicled concern the partition function \eqref{causalpartitionfunction} not the path sum \eqref{causalpathsum}. What can we conclude from these results concerning the Lorentzian sector of this quantum theory of gravity? Unfortunately, the absence of an Osterwalder-Schrader-type theorem precludes drawing definite conclusions \cite{KO&RS1,KO&RS2}. Reversing the Wick rotation is moreover not a viable option in the context of numerical simulations. Accordingly, one currently interprets the results optimistically along the most obvious of lines: that the path sum \eqref{causalquantumstate} exhibits a corresponding phase structure with the ensemble average quantum geometry of the corresponding phase C dominated on large length scales by Lorentzian de Sitter spacetime. The spectral dimension is not well-defined for Lorentzian manifolds, so the translation of its dynamical reduction to the Lorentzian sector is not clear, but the literature contains several suggestions \cite{GC&AE&FS,SC2}. 

\emph{Outlook}---Employing the well-established methods of lattice quantum chromodynamics and invoking just one key hypothesis, 
the causal dynamical triangulation approach has produced several promising results. 
In its most studied case, on which I concentrated above, the approach yields a classical limit, including semiclassical corrections, evidently consistent with general relativity, 
a phase diagram containing a second order transition, and the novel prediction of the dynamical reduction of dimension on sufficiently small length scales.

The first of these results partly addresses the first key question facing the causal dynamical triangulations approach: the recovery of the correct classical limit. In addition to subjecting the existing findings to further scrutiny and testing their validity in other cases, one should attempt to determine whether or not the approach yields the correct classical limit in the nonrelativistic regime. In other words, does Newtonian gravity emerge in an appropriate limit? 
Arguably, this is the most important test of any theory of gravity. 

The second of these results motivates the investigation of 
the second key question facing the causal dynamical triangulations approach: the existence of a continuum limit. The renormalization group analysis of \cite{JA&AG&JJ&AK&RL} bore equivocal results, owing partly to the computational challenges of running Markov chain Monte Carlo simulations near second order phase transitions. In addition to surmounting these technical issues, one should entertain alternative renormalization group schemes such as that proposed in \cite{JHC1}. 
Furthermore, since a renormalization group analysis is completely parasitic on one's knowledge of phenomenology, 
one should firmly press ahead with all manners of explorations. 


The third of these results 
reminds us that all of our physical understanding of the causal dynamical triangulation approach concerns its Euclidean sector. Supposing that, on the basis of Markov chain Monte Carlo simulations, one manages to answer in the affirmative either of the two key questions, how confident could one be that such a result would survive translation into the Lorentzian sector? With the presence of the distinguished foliation and the reflection positivity of the transfer matrix providing glimmers of hope 
for an Osterwalder-Schrader-type theorem, one should actively pursue such a result. Progress on this issue would bring us back to the motivating goal of making sense of the Lorentzian path integral \eqref{quantumstate}. 

Undoubtedly, there remain difficult problems to solve, 
but the questions on which causal dynamical triangulations either stands or falls are now within sight of being answered. 

\emph{Acknowledgments}---I acknowledge support from the Foundation for Fundamental Research on Matter itself supported by the Netherlands Organization for Scientific Research. 





\begin{thebibliography}{99}

\bibitem{LR1} L. Rosenfeld. ``Zur Quantelung der Wellenfelder." \emph{Annalen der Physik} 5 (1930) 113. 

\bibitem{LR2} L. Rosenfeld. ``Uber die Gravitationswirkungen des Lichtes." \emph{Zeitschrift fur Physik} 65 (1930) 589.

\bibitem{MPB} M. P. Bronstein. ``Quantentheories schwacher Gravitationsfelder." \emph{Physikalische Zeitschrift der Sowietunion} 9 (1936) 140.

\bibitem{GtH&MV}  G. Õt Hooft and M. Veltman. ``One-loop divergences in the theory of gravitation."\emph{Annales de l'Institut Henri Poincar\'e: Section A Physique Th\'eorique} 20 (1974) 69.

\bibitem{MHG&AS} M. H. Goroff and A. Sagnotti. ``The ultraviolet behavior of Einstein gravity." \emph{Nuclear Physics B} 266 (1986) 709.

\bibitem{JD} J. Donoghue. ``The effective field theory treatment of quantum gravity." arXiv: gr-qc/1209.3511.

\bibitem{SW} S. Weinberg. ``Ultraviolet divergences in quantum theories of gravitation." \emph{General Relativity: An Einstein Centenary Survey}. Eds. S. W. Hawking and W. Israel. Cambridge University Press, Cambridge, 1979.

\bibitem{BHW&DK&AW} B. H. Wellegehausen, D. K\"orner, and A. Wipf. ``Asymptotic safety on the lattice: the nonlinear $\mathrm{O}(N)$ sigma model." arXiv: hep-lat/1402.1851.


\bibitem{MN&MR} M. Niedermaier and M. Reuter. ``The asymptotic safety scenario in quantum gravity." \emph{Living Reviews in Relativity} 9 (2006) 5.

\bibitem{PH1} P. Ho\v{r}ava. ``Quantum gravity at a Lifshitz point." \emph{Physical Review D} 79 (2009) 084008.

\bibitem{SC} S. Carlip. ``Quantum gravity: a progress report." \emph{Reports on Progress in Physics} 64 (2001) 885.

\bibitem{QCD} Particle Data Group. ``Review of Particle Physics." \emph{Physical Review D} 86 (2012) 010001.

\bibitem{JA&JJ} J. Ambj\o rn and J. Jurkiewicz. ``Scaling in four-dimensional quantum gravity." \emph{Nuclear Physics B} 451 (1995) 643.

\bibitem{RG} R. Geroch. ``Topology in General Relativity." \emph{Journal of Mathematical Physics} 8 (1967) 782.

\bibitem{TR} T. Regge. ``General Relativity without Coordinates." \emph{Nuovo Cimento} 19 (1961) 558.

\bibitem{JA&JJ&RL2} J. Ambj{\o}rn, J. Jurkiewicz, and R. Loll. ``Dynamically triangulating Lorentzian quantum gravity." \emph{Nuclear Physics B} 610 (2001) 347.

\bibitem{JA&AG&JJ&RL3}  J. Ambj{\o}rn, A. G\"{o}rlich, J. Jurkiewicz, and R. Loll. ``Nonperturbative quantum gravity." \emph{Physics Reports} 519 (2012) 127.

\bibitem{KNA&TA&JN} K. N. Anagnostopoulos, T. Azuma, and J. Nishimura. ``Towards an effective importance sampling in Monte Carlo simulations of a system with a complex action." \emph{Proceedings of Science} Lattice 2011 (2011) 181. 

\bibitem{RL} R. Loll. ``Discrete Approaches to Quantum Gravity in Four Dimensions." \emph{Living Reviews in Relativity} 1 (1998) 13. http://www.livingreviews.org/lrr-1998-13.

\bibitem{JA&RL} J. Ambj\o rn and R. Loll. ``Non-perturbative Lorentzian quantum gravity, causality, and topology change." \emph{Nuclear Physics B} 536 (1998) 407.

\bibitem{ACJL}  J. Ambj{\o}rn, J. Correia, C. Kristjansen, and R. Loll. ``On the relation between Euclidean and Lorentzian 2D quantum gravity.'' \emph{Physics Letters B} 475 (2000) 24.

\bibitem{SJ&SDM} S. Jain and S. D. Mathur. ``World-sheet geometry and baby universes in 2D quantum gravity." \emph{Physics Letters B} 286 (1992) 239.

\bibitem{JA&JJ&RL1} J.Ambj{\o}rn, J. Jurkiewicz, and R. Loll. ``Non-perturbative Lorentzian Path Integral for Gravity." \emph{Physical Review Letters} 85 (2000) 347.

\bibitem{DM} D. Mattingly. ``Modern Tests of Lorentz Invariance." \emph{Living Reviews in Relativity} 8 (2005) 5.

\bibitem{JA&LG&YS&YW} J. Ambj\o rn, L. Glaser, Y. Sato, and Y. Watabiki. ``2d CDT is 2d Ho\v{r}ava-Lifshitz quantum gravity." \emph{Physics Letters B} 722 (22013) 172.

\bibitem{CDTandHL} J. Ambj{\o}rn, A. G\"{o}rlich, S. Jordan, J. Jurkiewicz, and R. Loll. ``CDT meets Ho\v{r}ava-Lifshitz gravity." \emph{Physics Letters B} 690 (2010) 413.

\bibitem{SJ&RL1} S. Jordan and R. Loll. ``Causal Dynamical Triangulations without preferred foliation." \emph{Physics Letters B} 724 (2013) 155. 

\bibitem{SJ&RL2} S. Jordan and R. Loll. ``De Sitter universe from causal dynamical triangulations with preferred foliation." \emph{Physical Review D} 88 (2013) 044055.

\bibitem{JA&SJ&JJ&RL1} J. Ambj{\o}rn, S. Jordan, J. Jurkiewicz, and R. Loll. ``Second-Order Phase Transition in Causal Dynamical Triangulations." \emph{Physical Review Letters} 107 (2011) 211303.

\bibitem{JA&SJ&JJ&RL2} J. Ambj{\o}rn, S. Jordan, J. Jurkiewicz, and R. Loll. ``Second- and first-order phase transitions in causal dynamical triangulations." \emph{Physical Review D} 85 (2012) 124044.

\bibitem{JA&AG&JJ&AK&RL} J. Ambj{\o}rn, A. G\"{o}rlich, J. Jurkiewicz, A. Kreienbuehl, and R. Loll. ``Renormalization group flow in CDT." \emph{Classical and Quantum Gravity} 31 (2014) 165003. 

\bibitem{JHC1} J. H. Cooperman. ``Renormalization of lattice-regularized quantum gravity models II. The case of causal dynamical triangulations." arXiv: gr-qc/1406.4531.

\bibitem{JA&JJ&RL4} J. Ambj{\o}rn, J. Jurkiewicz, and R. Loll. ``Emergence of a 4D World from Causal Dynamical Triangulations." \emph{Physical Review Letters} 93 (2004) 131301.

\bibitem{JA&JJ&RL5} J. Ambj{\o}rn, J. Jurkiewicz, and R. Loll. ``Semiclassical universe from first principles." \emph{Physics Letters B} 607 (2005) 205.

\bibitem{JA&JJ&RL7} J. Ambj{\o}rn, J. Jurkiewicz, and R. Loll. ``The Spectral Dimension of the Universe is Scale Dependent."  \emph{Physical Review Letters} 95 (2005) 171301.

\bibitem{JA&JJ&RL6} J. Ambj{\o}rn, J. Jurkiewicz, and R. Loll. ``Reconstructing the universe." \emph{Physical Review D} 72 (2005) 064014.

\bibitem{RK} R. K. Kommu. ``A validation of causal dynamical triangulations." \emph{Classical and Quantum Gravity} 29 (2012) 105003.

\bibitem{JA&AG&JJ&RL1} J. Ambj{\o}rn, A. G\"{o}rlich, J. Jurkiewicz, and R. Loll. ``Planckian Birth of a Quantum de Sitter Universe." \emph{Physical Review Letters} 100 (2008) 091304.

\bibitem{JA&AG&JJ&RL2}  J. Ambj{\o}rn, A. G\"{o}rlich, J. Jurkiewicz, and R. Loll. ``Nonperturbative quantum de Sitter universe."  \emph{Physical Review D} 78 (2008) 063544.

\bibitem{JA&AG&JJ&RL&JGS&TT} J. Ambj{\o}rn, A. G\"{o}rlich, J. Jurkiewicz, R. Loll, J. Gizbert-Studnicki, and T. Trze\'{s}niewski. ``The semiclassical limit of causal dynamical triangulations." \emph{Nuclear Physics B} 849 (2011) 144.

\bibitem{JA&JGS&AG&JJ} J. Ambj{\o}rn, J. Gizbert-Studnicki, A. G\"{o}rlich, and J. Jurkiewicz. ``The transfer-matrix in four-dimensional CDT." \emph{Journal of High Energy Physics} 2012 (2012) 17. 

\bibitem{DB&JH} D. Benedetti and J. Henson.  ``Spectral geometry as a probe of quantum spacetime." \emph{Physical Review D} 80 (2009) 124036.

\bibitem{JA&JJ&RL3} J. Ambj{\o}rn, J. Jurkiewicz, and R. Loll. ``Nonperturbative 3d Lorentzian Quantum Gravity." \emph{Physical Review D} 64 (2001) 044011.

\bibitem{JHC&JMM} J. H. Cooperman and J. M. Miller. ``A first look at transition amplitudes in $(2+1)$-dimensional causal dynamical triangulations." \emph{Classical and Quantum Gravity} (2014). 

\bibitem{JHC&KL&JMM} J. H. Cooperman, K. Lee, and J. M. Miller. ``A closer look at transition amplitudes in $(2+1)$-dimensional causal dynamical triangulations." In preparation.

\bibitem{TB&RL} T. Budd and R. Loll. ``Exploring torus universes with causal dynamical triangulations." \emph{Physical Review D} 88 (2013) 024015.

\bibitem{MKS} M. K. Sachs. ``Testing Lattice Quantum Gravity in $2+1$ Dimensions." arXiv: gr-qc/1110.6880.

\bibitem{JHC2} J. H. Cooperman. ``Scale-dependent homogeneity measures for causal dynamical triangulations." arXiv: gr-qc/1410.XXXX.

\bibitem{CA&SJC&JHC&PH&RKK&PZ} C. Anderson, S. J. Carlip, J. H. Cooperman, P. Ho\v{r}ava, R. K. Kommu, and P. R. Zulkowski. ``Quantizing Ho\v{r}ava-Lifshitz gravity via causal dynamical triangulations." \emph{Physical Review D} 85 (2012) 044027. 

\bibitem{KO&RS1} K. Osterwalder and R. Schrader. ``Axioms for Euclidean Green's functions." \emph{Communications in Mathematical Physics} 31 (1973) 83.

\bibitem{KO&RS2} K. Osterwalder and R. Schrader. ``Axioms for Euclidean Green's functions II." \emph{Communications in Mathematical Physics} 42 (1975) 281.

\bibitem{GC&AE&FS} G. Calcagni, A. Eichhorn, and F. Saueressig. ``Probing the quantum nature of spacetime by diffusion." \emph{Physical Review D} 87 (2013) 124028.

\bibitem{SC2} S. Carlip. ``The Small Scale Structure of Spacetime." \emph{Foundations of Space and Time}. Eds. G. Ellis, J. Murugan, and A. Weltman. Cambridge University Press 2010.

\end{thebibliography}
\end{document}